\documentclass{IEEEtran}

\usepackage{graphicx}
\usepackage{array}
\usepackage{cite}
\usepackage{url}
\usepackage[table]{xcolor}
\usepackage[dvipsnames]{xcolor}
\usepackage{hyperref}
\usepackage{amsmath,amsfonts}

\hypersetup{colorlinks=true, citecolor=blue,linkcolor=blue,filecolor=blue,urlcolor=blue}

\begin{document}

\title{Quantum Network Simulation and Emulation: A Roadmap for Quantum Internet Design}


\author{
\IEEEauthorblockN{Brian Doolittle and Michael Cubeddu} \\
\IEEEauthorblockA{\textit{Aliro Technologies, Inc.} \\ Brighton, MA, USA \\
Email: \{bdoolittle,mcubeddu\}@aliroquantum.com}}

\onecolumn{
\vspace*{\fill}
\begin{center}
    \Huge{This article has been accepted for publication in the IEEE
International Conference on Quantum Communications,
Networking, and Computing (QCNC 2026). This is the accepted
manuscript made available via arXiv.}
\end{center}

\vfill

\normalsize{\noindent© 2026 IEEE. Personal use of this material is permitted. Permission from IEEE must be obtained for all other uses, in any
current or future media, including reprinting/republishing this material for advertising or promotional purposes, creating new
collective works, for resale or redistribution to servers or lists, or reuse of any copyrighted component of this work in other
works.}

}
\newpage

\twocolumn

\maketitle

\begin{abstract}
Quantum networks are advancing the information technology infrastructure of society.
Simulation and emulation software tools have emerged to support the design, development, and deployment of quantum networks, however, classical simulation and emulation methods have major bottlenecks in the error, latency, and cost that they can achieve at scale.
In this work, we review quantum network simulation and emulation tools, including foundational principles, state-of-the-art tools, and bottlenecks.
We then discuss how quantum technologies can address these challenges, and we construct a roadmap for the adoption of quantum simulation and emulation tools, emphasizing  codesign with quantum network testbeds.
\end{abstract}


\IEEEpeerreviewmaketitle

\section{Introduction}

Quantum networks are rapidly being developed, deployed, and scaled, advancing applications in secure communications, distributed sensing, and distributed quantum computing \cite{wehner2018_quantum_internet}.
As quantum networks scale, their benefit to society will increase, but so will the demands placed on their design tools. The standard tools of quantum networking design include analytical methods and theoretical frameworks, information theoretic metrics and benchmarks,  and software tools for simulation, emulation, and optimization \cite{azuma2021_qn_design_tools}. We focus on software tools because they are constrained by the limitations of current information processing technology.

Historically, software-based design tools have addressed critical design challenges at scale in classical networking and communications. Simulation tools such as computer-aided design frameworks and emulation tools such as digital twins \cite{qi2021_digitial_twins}, have accelerated research and development across a wide range of industries. One of the key features of digital twins and network emulation tools, such as the Common Open Research Emulator (CORE) \cite{ahrenholz2008_core}, is their ability to test networking applications in a virtual environment while also being able to integrate with external, real-world software and devices.

It is no surprise that software-based design tools for simulation and emulation are already integral to the design and development of quantum networks \cite{abreu2024_simulator_comparison,bel2024_qn_sim_review,Hayek2026_qn_design_tools}. These tools include simulation software such as discrete event simulators and quantum circuit simulators, as well as, emulation software such as digital twins and virtual quantum networks.
Existing quantum network simulators have already demonstrated their ability to accelerate, validate, and benchmark designs for quantum network applications and protocols \cite{kozlowski2020_protocol_designing,chalupnik2025_realistic_bbm92_simulation,bhatia2025_qdba,wallnofer2022simulating_qkd_satellites,mehic2024_qkd_emulation,martin2024_qkd_service,diaz2025_qkd_digital_twin,yang2025_qkd_digital_twin,chiti2025_digital_twin_sdn_qn}.

Classical software is straightforward to develop and scale, but the fact remains that classical computers cannot generally simulate large quantum systems while maintaining low error, latency, and cost.
Indeed, the main bottlenecks of simulating quantum computers using classical processors and approximation methods are \cite{xu2025herculean}:

\begin{enumerate}
    \item \textbf{Scalability:} Classical processors cannot store large quantum states in memory, requiring approximation.
    \item  \textbf{Accuracy:} The approximations used on classical processors translates into simulation error.
    \item \textbf{Efficiency:} Classical processors require vast amounts of hardware and energy to model quantum systems with accuracy at scale.
    \item \textbf{Validity:} The expected behavior of a quantum system might not be known, making it difficult to verify that the simulation output is accurate.
\end{enumerate}

When simulating or emulating quantum networks on classical processors, these bottlenecks will lead to roadblocks that will prevent design tools from scaling. 
As quantum networks scale, there will be an inflection point where classical design tools fail to meet the  needs of industrial quantum networks.
However, the bottlenecks of simulating quantum networks can be addressed using quantum hardware to simulate and emulate large-scale quantum networks.
This work outlines a roadmap for quantum network design tools that begins with the state-of-the-art, and extrapolates to the tools necessary for quantum internet design. We highlight key challenge areas and the importance for codesign between computing and networking hardware.

This work is organized as follows: 
In Section~\ref{section:sim_vs_emu}, we formalize the foundational concepts, technical definitions, and performance metrics for quantum network simulation and emulation tools. In Section~\ref{section:cad_for_qn_soa}, we review state-of-the-art software-based design tools for quantum networks. In Section~\ref{section:roadmap}, we present a roadmap for development of quantum network simulation and emulation tools, discussing the feasibility and challenges of scaling classical software-based design tools, and how bottlenecks can be addressed by quantum technologies and codesign with hardware.
In Section~\ref{section:discussion}, we discuss the key directions in which software-based design tools must be developed to facilitate the broader development of the quantum internet.

\begin{figure*}
    \centering
    \includegraphics[width=\textwidth]{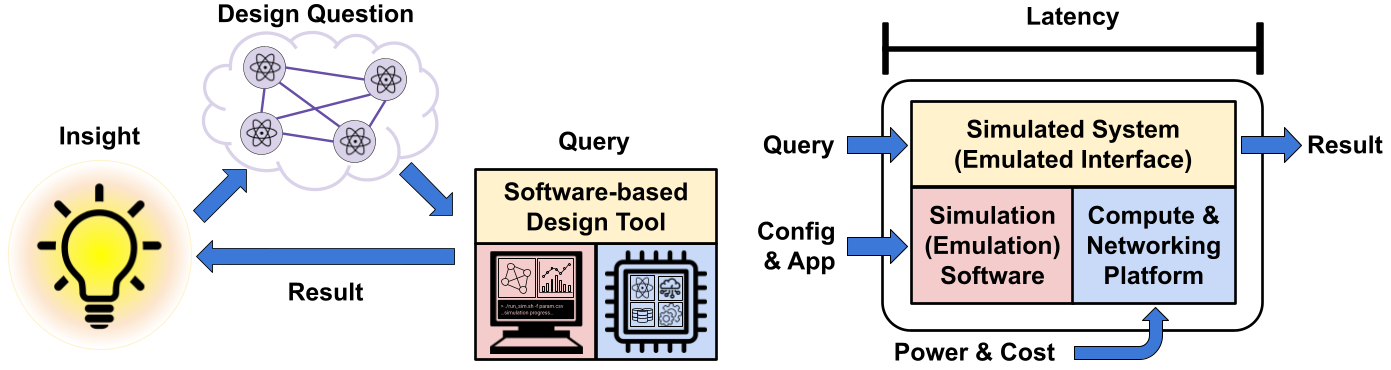}
    \caption{(Left) A design cycle in which a quantum networking design question is used to query a software-based design tool, the output data is used to gain insights, leading to development. New problems are encountered and the process iterates. (Right) A depiction of the key components of simulation or emulation tools, their inputs, and associated metrics.}
    \label{fig:design_process}
\end{figure*}

\section{Quantum Network Simulation and Emulation}\label{section:sim_vs_emu}

\subsection{Simulation-based Design vs. Emulation-based Design}

Simulation and emulation are both software-based design tools for quantum networks, but they are each used for distinct tasks.
Simulators estimate a physical system's properties, such as quantum state, probability distributions, or metrics, and are typically used as a predictive tool in research and design. Emulators replicate the real-time behavior of system interfaces and are used for engineering and testing networking applications and device integrations.
In general, simulation is a broader concept that includes emulation.

When using simulation or emulation tools, the quantum network design process is generally iterative (see Fig.~\ref{fig:design_process}). If simulation and emulation tools can operate with enough speed and a low enough cost, then they can accelerate the development of real-world systems. These software-based design tools allow the parameter of quantum network designs to be fine-tuned where numerous optimization frameworks and tools for quantum networks have been developed  \cite{doolittle2023_vqo,doolittle2024_operational_framework,avis2025_repeater_placement}.
In both simulation and emulation, the main advantage is that design and development can be performed rapidly, and independently from hardware acquisition.

\subsection{Simulation Tools vs. Emulation Tools}

Simulation and emulation are closely related concepts that each address key challenges in the design and development of quantum networks.
Simulators and emulators both model quantum network dynamics as a function, $y = f_{\texttt{Sim}}(x)$, in which a simulation query $x$ is transformed into a simulation result $y$.
The input $x$ contains the simulation parameters while the output $y$ contains the computed data and metrics.
Simulation and emulation tools may also be configured with contextual data, including the application, network topology, and simulation hyperparameters. Likewise, the compute platform running the simulation or emulation tool will have an associated power or cost that must also be  (see Fig.~\ref{fig:design_process}). 

The key distinction between simulation and emulation is that emulation reproduces the real-time behavior of a system's interface, whereas simulation is used to predict a system's behavior, properties, and metrics without any constraint on real-time operation. 
Since emulation aims to replicate the input-output behavior of a system in real-time, emulation tools can build upon the basic simulation workflow, making it interactive, operate in real-time, and able to integrate with other emulated or real-world networking devices (see Fig.~\ref{fig:emulation_interfaces}).

Since simulation tools do not need to  adhere to real-time constraints or real-world interfaces, they can be more flexible in their simulation queries and results. Simulation tools are generally simpler than emulation tools because real-time interactive behaviors do not have to be replicated. As a result, each simulation job can be treated independently, leading to simpler implementations of simulation tools. On the contrary, emulation tools require more complicated logic, including interactive APIs, stateful data management and context, and real-time performance.  

\subsection{Real-time Frameworks vs. Virtual-time Frameworks}

Emulations operate in \textit{real-time}, meaning they rely on the compute platform's clock time to model the timing characteristics of the emulated interface.  On the contrary, simulations can operate in \textit{virtual-time}, meaning the simulation has a virtual timeline that is independent from the compute platform's clock. 
Real-time frameworks (emulators) are valuable because they enable integration between emulated and real-world devices, as well as, application or control layer software to be ported without modification between an emulated physical layer and real-world quantum devices. 
Virtual-time simulation frameworks are valuable because they are not limited by real-world clocks, allowing high-resolution simulations of subnanosecond timescales, or simulations of timescales that exceed the design timeline.


\subsection{Realistic Modeling vs. Virtual Modeling}

Simulation and emulation both aim to produce accurate results, but this does not imply that the underlying data representation and compute platform resemble physical reality. 
Realism versus virtualism in physical modeling is a spectrum that qualifies the amount of detail to which  a model describes the inner workings of a physical system. The amount of realism in a simulation or emulation  can have major ramifications on its efficiency and accuracy. 
Realistic modeling accounts for each physical parameter and coupling in the system and aims to reproduce physical behavior from the ground up. Virtual modeling abstracts away low-level physical details and replaces them with simpler high-level models and representations. Note that physical models in classical software are represented by numeric data structures, which is a virtual representation that is distinct from physical reality.

The line between realism and virtualism in quantum network simulation and emulation tools is determined by the level of abstraction at which they operate. We summarize these levels of abstraction as three layers that resemble a networking software stack: application, control, and physical.
Higher levels of abstractions operate with greater latency than lower levels of abstraction, making it difficult to simulate or emulate all layers at once. As observed in Reference~\cite{ang2024arquin}, efficient simulation of a distributed quantum computing stack requires each layer to be simulated independently where the output of one layer's simulation informs the input of the next. As a result, quantum network emulation tools, including digital twins and virtual quantum networks, typically focus on a layer of abstraction at the expense of lower level modeling (see Section~\ref{section:digital_twins_and_virtual_quantum_networks}).



\subsection{Performance Metrics: Error, Latency, and Cost}

The performance of simulation and emulation tools can be measured in terms of their error, latency, and cost,
enabling direct performance comparisons between different tools for quantum network simulation and emulation.

The error of a simulation or emulation captures the accuracy of its results as distance between the expected behavior or metric $y\in \mathbb{R}^d $ and the mean simulated metric  $\bar{y}\in\mathbb{R}^d$. The mean error or accuracy can be calculated as a simple Euclidean distance
\begin{equation}\label{eq:result_norm}
    E = ||y - \bar{y}|| = \sqrt{\sum_{i} (y_i-\bar{y}_i)^2},
\end{equation}
or by another type of norm or comparison metric \cite{bel2024_qn_sim_review}.
No matter the norm, the accuracy of the simulation characterizes the error $E$ on average. Precision can also be considered as the standard deviation of the average error $E$.

The latency of a simulation or emulation describes the time elapsed between query and result,
\begin{equation}\label{eq:sim_emu_rate}
    T = \frac{1}{R} \cdot \frac{\texttt{seconds}}{\texttt{result}}, \quad\text{and}\quad  R = \frac{N}{\Delta t} \cdot\frac{\texttt{results}}{\texttt{second}}
\end{equation}
where $R$ is the rate at which simulation results can be output.
Here $N$ represents the number of results and $\Delta t$ is the duration of time to obtain the results. Emulation requires low latencies (high rates) to maintain real-time operation.

The cost provides a practical measure of the computational resources needed for the simulation task in terms of either power usage or monetary cost $\Omega \frac{\texttt{\$ or Joules}}{\texttt{result}}$. We define a simulation's total cost, $C$, and burn rate, $C_S$, as 
\begin{equation}\label{eq:sim_emu_cost}
    C = N \Omega \cdot (\texttt{\$}\text{ or }\texttt{Joules}) , \ \  C_S =  R\Omega \cdot(\frac{\texttt{\$}}{\texttt{s}}\text{ or }\texttt{Watts}).
\end{equation}
A monetary unit of cost is practical and can incorporate costs associated with power usage, labor, operations, maintenance, and hardware acquisition, whereas power usage offers a physical measure of simulation and emulation efficiency.

\subsection{Practicality of Simulation vs. Emulation}

For quantum network simulation or emulation tools to be practical, they must operate within a given project's constraints, including error tolerance, simulation latency, and budget. These performance constraints, $(\varepsilon, \tau, c)$, should be tailored to each project and are formalized by the following bounds:
\begin{enumerate}
    \item \textbf{Error:} The mean error, $E$, in Eq.~\eqref{eq:result_norm} is bound by the error tolerance, $\varepsilon$, as $E \leq \varepsilon$.
    \item \textbf{Latency:} The mean latency, $T$, from Eq.~\eqref{eq:sim_emu_rate} is bound by the latency tolerance, $\tau$, as $T = 1/R \leq \tau$.
    \item \textbf{Cost:} The total cost, $C$, from Eq.~\eqref{eq:sim_emu_cost} is bound by the budget, $c$, as $C\leq c$.
\end{enumerate}

For a simulation and emulation tool to be practical in a given project, each of these bounds should be satisfied. In general,  design tools should minimize the error, latency, and cost $(E, T, C)$ of their operation. 
However, tradeoffs can occur when optimizing these performance metrics \cite{xu2025herculean}. For instance, decreasing the error of a simulation  tool, typically requires extra computation, increasing the processing time and cost.
Conversely, lower latencies and lower costs can often be achieved by relaxing models, however, approximations can increase error.
Note that the real-time requirements of emulation tools place stringent requirements on the latency of their performance, which  inhibits the scale and accuracy of quantum network emulations.


\subsection{Validation and Verification of Simulation and Emulation}

It is important that simulation and emulation tools output accurate results. To validate the accuracy of a simulation or emulation tools, simulations must be benchmarked against established problems with behaviors that are either experimentally demonstrated or theoretically characterized. For example, in a recent work, simulations of an entanglement-based key distribution system are benchmarked against theoretical models and an experimental demonstration  \cite{chalupnik2025_realistic_bbm92_simulation}.
Validating and verifying simulation tools that are not experimentally or theoretically tractable, can be addressed by cross validation between different quantum network simulation and emulation tools. 
One example of cross validation of quantum network software includes benchmark comparisons between QuISP and SeQUeNCe \cite{chung2025_cross_validation}.

\section{State-of-the-Art Quantum Network Simulation and Emulation}\label{section:cad_for_qn_soa}

We now review the state-of-the-art tools for quantum network simulation and emulation.
For a broader comparison and benchmarking of the state-of-the-art, we refer the reader to references
\cite{abreu2024_simulator_comparison, bel2024_qn_sim_review, Hayek2026_qn_design_tools}.

\subsection{Physical Modeling of Quantum Network Devices}\label{section:physical_modeling}

Low-level details of network devices, components, and links are often simulated in isolation to characterize their input-output behavior. 
For example the physics of quantum memories, transducers, or sources can be simulated \cite{d2024_transducer_modeling,richardson2025_quantum_savory,robertson2025_memory_modeling} where the results are translated into efficiencies, fidelities, and operations that inform the higher-levels of abstraction in the simulation or emulation. Thus, low-level details can often be simulated separately from higher-level quantum network simulations. 

\subsection{Discrete Event Simulators for Quantum Networks}\label{section:des}

Discrete event simulation (DES) is a common approach to simulating multiagent scenarios. Several DES tools have been developed for quantum network design including NetSquid \cite{coopmans2021_netsquid}, SeQUeNCe \cite{wu2021_sequence}, QuISP \cite{2022quisp}, SimQN \cite{chen2023_simqn}, QuantumSavory.jl \cite{krastanov_quantumsavory,richardson2025_quantum_savory}, and Aliro Quantum Network Simulator \cite{chalupnik2025_realistic_bbm92_simulation,bhatia2025_qdba}.  
DES characteristically applies a virtual timeline, and commonly applies realistic representations of event timing, devices, and telemetric data. The multiagent aspect of DES enables modular integration of network devices allowing generic quantum networking systems to rapidly be constructed and iterated, lending way to success in applications including, satellite-based repeater networks \cite{wallnofer2022simulating_qkd_satellites}, or protocol design, verification, and benchmarking \cite{kozlowski2020_protocol_designing,chalupnik2025_realistic_bbm92_simulation,bhatia2025_qdba}.

\subsection{Digital Twins and Virtual Quantum Networks}\label{section:digital_twins_and_virtual_quantum_networks}

Digital twins have emerged to emulate the low-level interfaces of quantum networking devices.  
Field Programmable Gate Arrays (FPGAs) are the choice tool for emulation because they operate at high rates ($\sim1$ nanosecond), are reprogrammable, and support parallelized and pipelined processing. For instance, FPGAs have been used for emulating quantum key distribution (QKD) systems  \cite{li2021_fpga_emulation} and quantum processors  \cite{khalid2004fpga,mahmud2019_scaling_qc_emulation,el2023_qc_classical_emulation}. Analog circuit-based approaches have also been developed for emulating quantum systems \cite{mourya2023_qc_emulation}.  
Virtual quantum networks have emerged as a high-level design tool for emulating the control and application layers of quantum networks. 
Virtual quantum networks, such as SimulaQron \cite{dahlberg2018_simulaqron} and QuNetSim \cite{diadamo2021_qunetsim}, can accelerate application and protocol development for quantum networks by providing a virtual testbed.
Virtual quantum networks have been applied to emulating quantum computing cloud infrastructure \cite{luo2025_digital_twin_scalable_cloud},  emulating QKD  \cite{mehic2024_qkd_emulation,martin2024_qkd_service,diaz2025_qkd_digital_twin,yang2025_qkd_digital_twin},  and emulating entanglement distribution in satellite-based networks \cite{chiti2025_digital_twin_sdn_qn}.
A drawback of virtual quantum networks and digital twins is that  they often ignore any realistic modeling of quantum systems to achieve real-time operation, however, they can be enhanced with lower-level modeling if processing power permits \cite{diadamo2021_integrating_quantum_simulation}.

\begin{figure}
    \centering
    \includegraphics[width=\columnwidth]{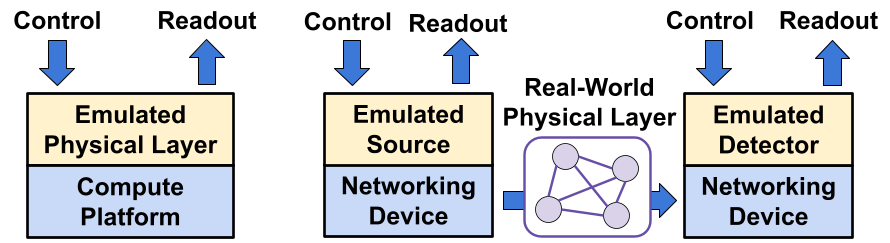}
    \caption{(Left) The physical layer of a network is emulated on a monolithic compute platform. (Right) An emulated source and detector connect over a real-world classical or quantum physical layer.}
    \label{fig:emulation_interfaces}
\end{figure}

\subsection{Quantum Simulation and Emulation of Quantum Networks}

Quantum processors present a natural platform for simulating quantum networks because they can efficiently simulate the quantum dynamics of highly entangled states. Quantum simulation is a key application of quantum computing \cite{lloyd1996universal,georgescu2014_quantum_simualation_review}, but its application in quantum network simulation and emulation did not emerge until later \cite{altman2021_quantum_sim}. Similarly, quantum twin emulation systems that make use of quantum processors and quantum interfaces (see Fig.~\ref{fig:emulation_interfaces} and~\ref{fig:quantum_emu.png}), are projected to have broad impact across many industries \cite{amir2022_quantum_twins}.

To simulate a quantum network on a quantum computer, the main approach is to map the quantum network's state preparations, device operations communication channels, noise, and measurements into a quantum circuit \cite{doolittle2023_vqo,doolittle2024_operational_framework,riera2025_qn_simulation}. If the quantum circuit is executed without noise, the stochastic behavior of the network is reproduced exactly in each shot. 
Quantum computers have also been applied to quantum network emulations by using larger quantum processors to emulate systems and networks of smaller quantum processors \cite{haner2016_qc_emulator, elyasi2025framework, kefaloukos2025internet}.
A quantum computer is expected to emulate at a higher rate than classical computers because it can evaluate the quantum circuits with greater efficiency.
Although these approaches show promise in simulating large scale networks, the noise and scale of existing quantum computers has not shown any advantage over classical processors \cite{doolittle2023_vqo,chen2023inferring_qn_topology,riera2025_qn_simulation}.

\section{A Roadmap for Quantum Internet Design }\label{section:roadmap}

\begin{figure}[t]
    \centering
    \includegraphics[width=\columnwidth]{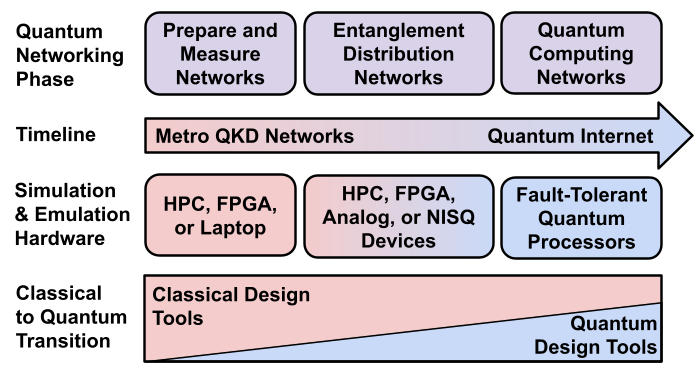}
    \caption{A roadmap for quantum network simulation and emulation tools is shown against a timeline ranging from today's metropolitan QKD networks to the future quantum internet. Three phases of capabilities are shown, prepare and measure networks, entanglement distribution networks, and quantum computing networks. The expected compute platforms for each phase are listed, where quantum hardware will gradually replace classical compute platforms for simulation and emulation of the physical layer. In the long-term, classical compute platforms are expected to support simulation and emulation of classical processes that occur in the control and application layers. }
    \label{fig:sim_emu_timeline}
\end{figure}

In this section, we present our roadmap for quantum network simulation and emulation tools (see Fig.~\ref{fig:sim_emu_timeline}).
As quantum networking technologies develop, their rates, scale, and complexity will increase, requiring simulation and emulation tools to accommodate more qubits and faster rates. These mounting challenges indicate that classical simulation and emulation tools will be insufficient for the design of future quantum networks. To build a quantum internet, we must transition from classical design tools to quantum design tools.
Furthermore, codesign between quantum networks and their software-based design tools will facilitate the iterative development of the quantum internet.

\subsection{Quantum Network Stages of Development}

Quantum networks are predicted to develop in stages of key capabilities \cite{wehner2018_quantum_internet}. We summarize these stages as follows:

\begin{enumerate}
    \item \textbf{Prepare and Measure (PM) Networks:} Photonic devices provide point-to-point quantum communication, enabling applications in secure communications over trusted relay nodes, and requiring high rates of point-to-point quantum communication.  
    \item \textbf{Entanglement Distribution (ED) Networks:} Quantum repeaters facilitate long-distance entanglement distribution, enabling applications in end-to-end secure communications and distributed quantum sensing and requiring quantum memories, transduction, and noisy intermediate scale quantum (NISQ) information processing. 
    \item \textbf{Quantum Computing (QC) Networks:} Entanglement is shared between quantum processors, enabling applications in distributed quantum computing and requiring fault-tolerant quantum processors with on-demand entanglement generation.
\end{enumerate}
In each stage, the requirements on fidelity and rates become more stringent, causing bottlenecks in quantum network simulation and emulation. Furthermore, the evolving landscape of technologies will require design tools to adapt accordingly.

\subsection{Scaling Classical Simulation of Quantum Networks}

In general, classical computers cannot simulate or emulate quantum systems with low latency and low error, however, a range of technical approaches can be used to navigate the roadblock. First, high-performance computing (HPC) techniques, including parallelism, pipelining, and hardware acceleration, can speedup computations of quantum operations \cite{doi2019_qc_sim_hpc}.
For instance, graphics processing units (GPUs) have shown advantages in simulating quantum systems with many qubits \cite{zhang2015_gpu,horii2023_gpu,faj2023_gpu} while FPGAs and analog circuits have shown low latency emulation of quantum systems \cite{khalid2004fpga,mahmud2019_scaling_qc_emulation,el2023_qc_classical_emulation,mourya2023_qc_emulation}.
Second, the algorithms and data representations have a significant effect on the performance.
For instance, the most accurate description of quantum systems is the density matrix representation, however, the latency of simulating density matrix systems increases exponentially with the number of qubits. On the contrary, if the stabilizer formalism is used, the simulation latency will scale polynomially with the number of qubits, but simulation error will increase because the stabilizer formalism does not accurately model certain types of noise and operations \cite{2022quisp,xu2025herculean}. Third, the level of abstraction in modeling quantum devices and network layers can have a major impact on performance. To speedup simulation it is important to abstract away as many low-level details as possible \cite{ang2024arquin}, however, doing so will increase the error.

The main reason that classical simulation of quantum networks is thought to be feasible is that bipartite entanglement can be simulated on classical processors with both low error and low latency. Since bipartite entanglement is the main resource in most PM and ED network protocols, classical simulation and emulation tools are expected to be practical for the design of many applications in PM and ED networks  \cite{kozlowski2020_protocol_designing,li2021_fpga_emulation,wallnofer2022simulating_qkd_satellites,mehic2024_qkd_emulation,martin2024_qkd_service,diaz2025_qkd_digital_twin,yang2025_qkd_digital_twin,chalupnik2025_realistic_bbm92_simulation,chiti2025_digital_twin_sdn_qn}.
For instance, classical simulation tools are capable of simulating ED networks with up to 100 nodes, provided that they use an efficient state representation, such as the stabilizer formalism or error basis \cite{2022quisp}.

Although classical approaches can already tackle many practical quantum network design tasks, the modeled systems are often idealized. As more realistic modeling is incorporated into the physical layer, scaling challenges will emerge when modeling the physical behaviors and real-time speeds of quantum networking devices.
Although PM networks have limited quantum information processing capabilities, one challenge of their emulation is achieving the subnanosecond latencies required to model device physics and optics. Furthermore, realistic modeling can rapidly become intractable on classical processors if the physical systems grows too large due to coupling with external systems and sources of noise.
These scaling issues will be exacerbated in ED networks where complex multiparty entanglement can be constructed and processed.
We expect that HPC or FPGAs will be sufficient to overcome these challenges in the near-term, but their scale will be limited, inhibiting our ability to design quantum networks.

Designing QC networks with classical software is somewhat paradoxical because QC networks have fault-tolerant quantum processors and on-demand entanglement generation. Since fault-tolerant quantum processors offer information processing advantages over classical processors, it is unlikely that classical hardware would be beneficial to simulating the physical layer of QC networks. Nevertheless, the quantum internet is expected to have classical application and control layers \cite{wehner2018_quantum_internet}. These layers will require classical emulation tools for their development and design, hence classical simulation and emulation tools will continue to play a role in quantum network design in the long-term.

\subsection{Quantum Simulation for Quantum Network Design}\label{section:quantum_sim_qn}

\begin{figure*}
    \centering
    \includegraphics[width=\textwidth]{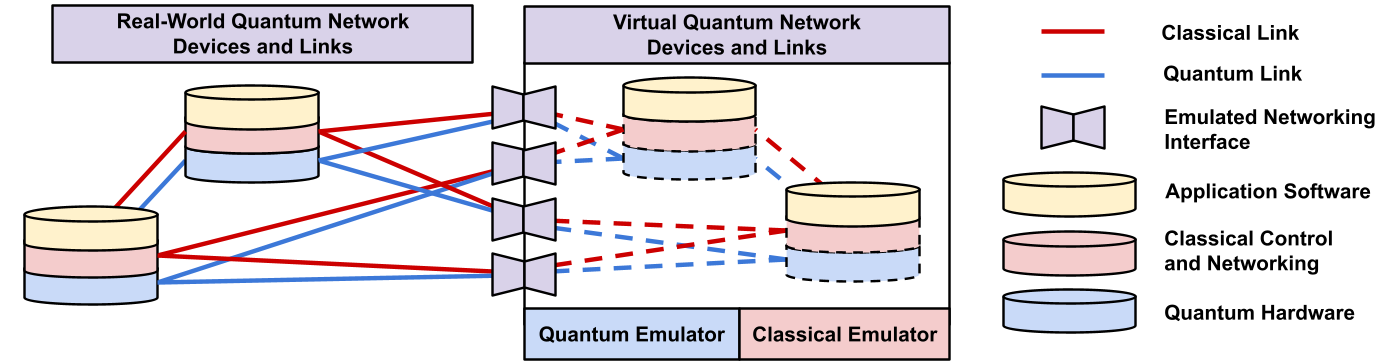}
    \caption{Hybrid scenario integrating real-world quantum network devices (solid lines) and virtual quantum network devices emulated on quantum processors (dashed lines). The application software layer (yellow) natively runs in both the real-world and virtual environments. Emulated quantum interfaces between real-world (left) and emulated devices (right) are achieved by quantum transduction while emulated classical interfaces are achieved by conventional networking hardware. Quantum hardware (blue) is emulated on quantum processors while classical software (red) is emulated on classical processors. }
    \label{fig:quantum_emu.png}
\end{figure*}

The advantage of using quantum hardware to simulate quantum networks is that the latency and accuracy of low-level modeling is improved, enabling detailed simulations of the physical layer.
However, initial demonstrations of quantum network simulation on NISQ processors show considerable amounts of error due to hardware noise \cite{doolittle2023_vqo,chen2023inferring_qn_topology,riera2025_qn_simulation}.
Nevertheless, simulations on NISQ devices can serve as a lower bound on real-world performance while quantum noise mitigation \cite{cai2023_error_mitigation} and error correction \cite{roffe2019_qec} techniques can reduce the simulation error further. To scale quantum network simulations on NISQ processors, circuit cutting techniques can be used to decompose large quantum circuits into smaller quantum circuits   \cite{peng2020_quantum_circuit_cutting,harada2024_doubly_optimal_wire_cutting}. These circuits can then be run at scale in quantum data centers, however, specialized infrastructure \cite{shapourian2025_quantum_datacenter} and large amounts of energy \cite{martin2022_quantum_datacenter} are needed. Therefore, the low latency and low error offered by quantum hardware might come at a significant cost. 

Although quantum computers can speed up simulations in all stages of quantum networking, the near-term challenge is that quantum hardware is too noisy \cite{Preskill2018_nisq}. NISQ processors need to demonstrate sufficiently low simulation cost, latency, and error in order to support practical quantum network design. Since NISQ processors have not demonstrated practical simulation for quantum networks, the near-term focus should be on scaling classical tools for PM and ED network design and developing frameworks for simulating quantum networks on quantum processors, such as \cite{doolittle2023_vqo,doolittle2024_operational_framework,kefaloukos2025internet,riera2025_qn_simulation}.
Once NISQ processors demonstrate advantages over classical processors, then these simulation frameworks can be used to simulate the physical layer and quantum information processing logic of ED and QC networks. 
As fault-tolerant quantum processors emerge, they will be a natural tool for QC network design.

\subsection{Quantum Emulation for Quantum Network Design}\label{section:quantum_emu_qn}

Since quantum network emulators replicate the real-time behavior of network interfaces, emulation tools will need to reproduce both classical and quantum interfaces. Quantum hardware will be necessary to provide emulation of quantum interfaces and to achieve low latency and error.
In  previous works \cite{haner2016_qc_emulator, elyasi2025framework, kefaloukos2025internet}, quantum networks are emulated on monolithic quantum processors, such that the physical layer of the network is emulated and its classical interface is exposed. In this approach, the quantum computer runs a circuit that encodes a realistic description of the quantum network devices, links, and operations. Since the quantum circuit reproduces the measurement statistics of the modeled quantum system, quantum computers are attractive for emulation tasks. Aside from noise, the main difficulty of emulation on quantum computers is the real-time control, operation, and readout of the quantum network interface. Furthermore, each quantum compute platform demonstrates a particular gate latency, qubit coherence time, and gate fidelity \cite{Ladd2010_qc,resch2019_cq_comparison}, care should be taken to select a platform with sufficiently low latency and sufficiently high gate fidelity and coherence time to maintain accurate emulation of quantum networks.

Quantum networks of all stages will require quantum twins \cite{amir2022_quantum_twins} to emulate quantum interfaces in the physical layer, enabling qubits or entanglement to be shared between real-world and virtual devices (see Fig.~\ref{fig:quantum_emu.png}). Integration between real-world and virtual devices facilitates robust testing of quantum network hardware and applications, and is a common feature of classical network emulation frameworks, such as CORE \cite{ahrenholz2008_core}.
In practice, integrations between real-world and virtual quantum networking devices over an emulated interface will require efficient  quantum transduction \cite{lauk2020_transduction} to translate between the qubit modality of the quantum emulation hardware and the quantum network hardware. Since efficient transduction will likely be needed for ED networks, the generalized emulation systems as shown in Fig.~\ref{fig:quantum_emu.png} will not begin to emerge until this stage.
In the near-term, quantum interfaces of PM devices could be emulated using tunable and reconfigurable optical sources or detectors supported by classical processing (see Fig.~\ref{fig:emulation_interfaces}). Such devices are the foundation of self-testing protocols \cite{Supic2020_self-testing}, in which an emulated quantum source or detector infers and certifies the capabilities, resources, and performance of a real-world quantum network device or link.

\begin{figure*}
    \centering
    \begin{tabular}{|m{1.4cm}|m{4cm}|m{6.5cm}|m{4.5cm}|}
        \hline
         \cellcolor{yellow!15}  & \cellcolor{yellow!15} \textbf{Simulation} & \cellcolor{yellow!15}  \textbf{Emulation (Classical) } & \cellcolor{yellow!15}  \textbf{Emulation (Quantum)}  \\
         \hline
        \textbf{Classical} \cellcolor{red!15} & \cellcolor{red!15} \footnotesize{Physical layer modeling for noisy PM and ideal ED networks. See Sections~\ref{section:physical_modeling} and~\ref{section:des}. } & \footnotesize{Physical layer emulation for PM networks or application/control layer emulation for ED/QC networks. See Section~\ref{section:digital_twins_and_virtual_quantum_networks}.} \cellcolor{red!15} & \footnotesize{Source and detector emulation, enabling device self-testing for PM networks. See Section~\ref{section:quantum_emu_qn}, Fig.~\ref{fig:emulation_interfaces}.} \cellcolor{red!15} \\
         \hline
         \textbf{Quantum} \cellcolor{RoyalBlue!15} & \footnotesize{Physical layer modeling for noisy ED/QC networks. See Section~\ref{section:quantum_sim_qn}} \cellcolor{RoyalBlue!15}  & \footnotesize{Physical layer emulation for ED/QC networks, enabling fullstack virtual quantum networks. See Section~\ref{section:quantum_emu_qn}.} \cellcolor{RoyalBlue!15}  & \footnotesize{Real and virtual quantum device integration. See Section~\ref{section:quantum_emu_qn}, Fig.~\ref{fig:quantum_emu.png}.} \cellcolor{RoyalBlue!15}  \\
        \hline
    \end{tabular}
    \caption{Expected applications of quantum network simulation and emulation of (classical/quantum) interfaces on both classical and quantum processors.}
    \label{table:sim_emu_taxonomy}
\end{figure*}

\subsection{Codesign of Quantum Networks and Design Tools}

It is paramount that simulation and emulation tools serve the needs of quantum network design, development, and deployment. In general, simulation and emulation tools should facilitate rapid and iterative development towards scaling the existing infrastructure while addressing the roadblocks that prevent the next stage of quantum networking. However, we should be cautious about developing tools that solve problems in the distant future because paradigms will shift, standards will changes, and technologies will become obsolete. Therefore, simulation and emulation tools should be developed in codesign with quantum network testbeds and industrial deployments.

Simulation tools can inform quantum network testbeds and vice versa. Indeed, experimental data from the testbeds can be used to create and validated models of real-world quantum networks. Simulation tools can then extrapolate these models to industrial scales and applications, informing the design, development, and deployment of quantum networking infrastructure. As the capabilities of quantum network testbeds improve, simulation and emulation tools will encounter bottlenecks. These bottlenecks will inform the key problems that simulation tools must address. Similarly, simulation tools can identify target parameters needed for testbeds to demonstrate for practical quantum network deployments.

Codesign of quantum networks and emulation tools is facilitated through standardized interfaces between network devices and layers of abstraction. A standardized classical interface enables direct comparison between emulated and real-world data, allowing quantum networking design tools to be compared and verified more easily. A standardized quantum interface enables self-testing workflows to be used to test the quantum resources and capabilities of quantum network devices and links.  
Thus, codesign of quantum network testbeds and quantum network emulators will lead to scalable and accurate tools for designing, developing, and deploying quantum networks via standardization.

\section{Discussion}\label{section:discussion}

In this work we analyze quantum network simulation and emulation tools while outlining a roadmap  for future development (see Fig.~\ref{fig:sim_emu_timeline}).  
Our main observation is that quantum network simulation and emulation each require low latency, low cost, and low error, where we expect that classical tools will encounter serious bottlenecks at scale. As summarized in Fig.~\ref{table:sim_emu_taxonomy}, classical and quantum processors will each play important role in the design and development of quantum networks where we expect that the transition from classical to quantum design tools will be among the first industrial demonstrations of quantum advantage. Finally, we emphasize the importance of codesign between quantum network testbeds and software-based design tools to ensure validation, standardization, and rapid transition from testbed to industrial application.
In this manner, quantum network simulation and emulation will facilitate and accelerate the design of the quantum internet \cite{wehner2018_quantum_internet}.

As a call to action, we highlight three key directions in which  quantum network design tools can be developed:
\begin{enumerate}
    \item Classical simulation and emulation tools for quantum networks must be scaled and their latency, error, and cost must be benchmarked for comparisons.
    \item Quantum simulation and emulation tools for quantum networks must be developed and benchmarked against classical tools to demonstrate practical application of quantum processors.
    \item  The classical and quantum interfaces between quantum network devices and layers of abstraction must be standardized and their simulation and emulation tools must be validated against quantum network testbeds. 
\end{enumerate}

\section*{Acknowledgments}

This work was funded by Aliro Technologies, Inc.

\bibliographystyle{ieeetr}
\bibliography{references}

@article{wehner2018_quantum_internet,
  title={Quantum internet: A vision for the road ahead},
  author={Wehner, Stephanie and Elkouss, David and Hanson, Ronald},
  journal={Science},
  volume={362},
  number={6412},
  pages={eaam9288},
  year={2018},
  publisher={American Association for the Advancement of Science}
}

@article{Preskill2018_nisq,
  doi = {10.22331/q-2018-08-06-79},
  url = {https://doi.org/10.22331/q-2018-08-06-79},
  title = {Quantum {C}omputing in the {NISQ} era and beyond},
  author = {Preskill, John},
  journal = {{Quantum}},
  issn = {2521-327X},
  publisher = {{Verein zur F{\"{o}}rderung des Open Access Publizierens in den Quantenwissenschaften}},
  volume = {2},
  pages = {79},
  month = aug,
  year = {2018}
}

@article{qi2021_digitial_twins,
  title={Enabling technologies and tools for digital twin},
  author={Qi, Qinglin and Tao, Fei and Hu, Tianliang and Anwer, Nabil and Liu, Ang and Wei, Yongli and Wang, Lihui and Nee, Andrew YC},
  journal={Journal of Manufacturing Systems},
  volume={58},
  pages={3--21},
  year={2021},
  publisher={Elsevier}
}

@inproceedings{ahrenholz2008_core,
  title={CORE: A real-time network emulator},
  author={Ahrenholz, Jeff and Danilov, Claudiu and Henderson, Thomas R and Kim, Jae H},
  booktitle={MILCOM 2008-2008 IEEE Military Communications Conference},
  pages={1--7},
  year={2008},
  organization={IEEE}
}

@article{azuma2021_qn_design_tools,
  title={Tools for quantum network design},
  author={Azuma, Koji and B{\"a}uml, Stefan and Coopmans, Tim and Elkouss, David and Li, Boxi},
  journal={AVS Quantum Science},
  volume={3},
  number={1},
  year={2021},
  publisher={AIP Publishing}
}

@inproceedings{abreu2024_simulator_comparison,
  title={Multipurpose quantum network simulators: A comparative study},
  author={Abreu, Diego and Pimentel, Arthur and Moraes, Polyana and Tavares, David and Veloso, Alan and Abel{\'e}m, Ant{\^o}nio},
  booktitle={Workshop de Redes Qu{\^a}nticas},
  pages={25--30},
  year={2024},
  organization={SBC}
}

@article{xu2025herculean,
  title={A herculean task: Classical simulation of quantum computers},
  author={Xu, Xiaosi and Benjamin, Simon and Chen, Jianxin and Sun, Jinzhao and Yuan, Xiao and Zhang, Pan},
  journal={Science Bulletin},
  year={2025},
  publisher={Elsevier}
}

@inproceedings{kozlowski2020_protocol_designing,
  title={Designing a quantum network protocol},
  author={Kozlowski, Wojciech and Dahlberg, Axel and Wehner, Stephanie},
  booktitle={Proceedings of the 16th international conference on emerging networking experiments and technologies},
  pages={1--16},
  year={2020}
}

@article{bhatia2025_qdba,
  title={$\backslash$textit $\{$In Silico$\}$ Benchmarking of Detectable Byzantine Agreement in Noisy Quantum Networks},
  author={Bhatia, Mayank and Doshi, Shaan and Winton, Daniel and Doolittle, Brian and Abreu, Bruno and N{\'u}{\~n}ez-Corrales, Santiago},
  journal={arXiv preprint arXiv:2509.02629},
  year={2025}
}

@article{wallnofer2022simulating_qkd_satellites,
  title={Simulating quantum repeater strategies for multiple satellites},
  author={Walln{\"o}fer, Julius and Hahn, Frederik and G{\"u}ndo{\u{g}}an, Mustafa and Sidhu, Jasminder S and Wiesner, Fabian and Walk, Nathan and Eisert, Jens and Wolters, Janik},
  journal={Communications Physics},
  volume={5},
  number={1},
  pages={169},
  year={2022},
  publisher={Nature Publishing Group UK London}
}

@article{robertson2025_memory_modeling,
  title={A digital twin of atomic ensemble quantum memories},
  author={Robertson, Elizabeth and Maa{\ss}, Benjamin and Tschernig, Konrad and Wolters, Janik},
  journal={arXiv preprint arXiv:2506.20403},
  year={2025}
}

@article{d2024_transducer_modeling,
  title={Simulation of quantum transduction strategies for quantum networks},
  author={d'Avossa, Laura and Zhan, Caitao and Chung, Joaquin and Kettimuthu, Rajkumar and Cacciapuoti, Angela Sara and Caleffi, Marcello},
  journal={arXiv preprint arXiv:2411.11377},
  year={2024}
}

@article{ang2024arquin,
  title={ARQUIN: architectures for multinode superconducting quantum computers},
  author={Ang, James and Carini, Gabriella and Chen, Yanzhu and Chuang, Isaac and Demarco, Michael and Economou, Sophia and Eickbusch, Alec and Faraon, Andrei and Fu, Kai-Mei and Girvin, Steven and others},
  journal={ACM Transactions on Quantum Computing},
  volume={5},
  number={3},
  pages={1--59},
  year={2024},
  publisher={ACM New York, NY}
}

@article{coopmans2021_netsquid,
  title={Netsquid, a network simulator for quantum information using discrete events},
  author={Coopmans, Tim and Knegjens, Robert and Dahlberg, Axel and Maier, David and Nijsten, Loek and de Oliveira Filho, Julio and Papendrecht, Martijn and Rabbie, Julian and Rozpedek, Filip and Skrzypczyk, Matthew and others},
  journal={Communications Physics},
  volume={4},
  number={1},
  pages={164},
  year={2021},
  publisher={Nature Publishing Group UK London}
}

@article{wu2021_sequence,
  title={SeQUeNCe: a customizable discrete-event simulator of quantum networks},
  author={Wu, Xiaoliang and Kolar, Alexander and Chung, Joaquin and Jin, Dong and Zhong, Tian and Kettimuthu, Rajkumar and Suchara, Martin},
  journal={Quantum Science and Technology},
  volume={6},
  number={4},
  pages={045027},
  year={2021},
  publisher={IOP Publishing}
}

@INPROCEEDINGS{2022quisp,
  author={Satoh, Ryosuke and Hajdušek, Michal and Benchasattabuse, Naphan and Nagayama, Shota and Teramoto, Kentaro and Matsuo, Takaaki and Metwalli, Sara Ayman and Pathumsoot, Poramet and Satoh, Takahiko and Suzuki, Shigeya and Meter, Rodney Van},
  booktitle={2022 IEEE International Conference on Quantum Computing and Engineering (QCE)}, 
  title={QuISP: a Quantum Internet Simulation Package}, 
  year={2022},
  volume={},
  number={},
  pages={353-364},
  doi={10.1109/QCE53715.2022.00056}}

@article{chen2023_simqn,
  title={SimQN: A network-layer simulator for the quantum network investigation},
  author={Chen, Lutong and Xue, Kaiping and Li, Jian and Yu, Nenghai and Li, Ruidong and Sun, Qibin and Lu, Jun},
  journal={IEEE Network},
  volume={37},
  number={5},
  pages={182--189},
  year={2023},
  publisher={IEEE}
}

@misc{krastanov_quantumsavory,
  author = {Krastanov, Stefan and et al.},
  title = {QuantumSavory.jl},
  year = {2023},
  publisher = {GitHub},
  journal = {GitHub repository},
  howpublished = {\url{https://github.com/QuantumSavory/QuantumSavory.jl}},
  commit = {4f57d6a0e4c030202a07a60bc1bb1ed1544bf679}
}

@article{richardson2025_quantum_savory,
  title={Full-stack Physics-level model of cascaded entanglement links},
  author={Richardson, J Gabriel and Dhara, Prajit and Bhatt, Abhishek and Guha, Saikat and Krastanov, Stefan},
  journal={arXiv preprint arXiv:2510.17976},
  year={2025}
}

@misc{bel2024_qn_sim_review,
      title={Simulators for Quantum Network Modelling: A Comprehensive Review}, 
      author={Oceane Bel and Mariam Kiran},
      year={2024},
      eprint={2408.11993},
      archivePrefix={arXiv},
      primaryClass={quant-ph},
      url={https://arxiv.org/abs/2408.11993}, 
}

@article{Hayek2026_qn_design_tools,
  title = {A Review of Software for Designing and Operating Quantum Networks},
  volume = {9},
  ISSN = {2511-9044},
  url = {http://dx.doi.org/10.1002/qute.202500808},
  DOI = {10.1002/qute.202500808},
  number = {2},
  journal = {Advanced Quantum Technologies},
  publisher = {Wiley},
  author = {Hayek,  Robert J. and Chung,  Joaquin and Kettimuthu,  Rajkumar},
  year = {2026},
  month = feb 
}

@article{dahlberg2018_simulaqron,
  title={SimulaQron-a simulator for developing quantum internet software},
  author={Dahlberg, Axel and Wehner, Stephanie},
  journal={Quantum Science and Technology},
  volume={4},
  number={1},
  pages={015001},
  year={2018},
  publisher={IOP Publishing}
}

@article{diadamo2021_qunetsim,
  title={Qunetsim: A software framework for quantum networks},
  author={DiAdamo, Stephen and N{\"o}tzel, Janis and Zanger, Benjamin and Be{\c{s}}e, Mehmet Mert},
  journal={IEEE Transactions on Quantum Engineering},
  volume={2},
  pages={1--12},
  year={2021},
  publisher={IEEE}
}

@article{diadamo2021_integrating_quantum_simulation,
  title={Integrating quantum simulation for quantum-enhanced classical network emulation},
  author={DiAdamo, Stephen and N{\"o}tzel, Janis and Sekav{\v{c}}nik, Simon and Bassoli, Riccardo and Ferrara, Roberto and Deppe, Christian and Fitzek, Frank HP and Boche, Holger},
  journal={IEEE Communications Letters},
  volume={25},
  number={12},
  pages={3922--3926},
  year={2021},
  publisher={IEEE}
}

@article{mehic2024_qkd_emulation,
  title={Emulation of quantum key distribution networks},
  author={Mehic, Miralem and Dervisevic, Emir and Burdiak, Patrik and Lipovac, Vlatko and Fazio, Peppino and Voznak, Miroslav},
  journal={Ieee network},
  volume={39},
  number={1},
  pages={116--123},
  year={2024},
  publisher={IEEE}
}

@article{martin2024_qkd_service,
  title={Service for deploying digital twins of qkd networks},
  author={Martin, Raul and Lopez, Blanca and Vidal, Ivan and Valera, Francisco and Nogales, Borja},
  journal={Applied Sciences},
  volume={14},
  number={3},
  pages={1018},
  year={2024},
  publisher={MDPI}
}

@inproceedings{diaz2025_qkd_digital_twin,
  title={A Digital Twin Approach to Quantum Key Distribution Under Eavesdropping},
  author={Diaz-Bricio, Angela and Lopez, Blanca and Vidal, Ivan and Valera, Francisco},
  booktitle={2025 International Conference on Quantum Communications, Networking, and Computing (QCNC)},
  pages={143--150},
  year={2025},
  organization={IEEE}
}

@inproceedings{yang2025_qkd_digital_twin,
  title={Demonstration of A Digital Twin for An Entanglement-Based Quantum Network},
  author={Yang, R and Wang, Rui and Clark, MJ and Tan, Jianxiong and Tai, J and Balram, Ronaldo and Jiang, Shan and Joshi, Siddarth Koduru and Simeonidou, Dimitra},
  booktitle={2025 Optical Fiber Communications Conference and Exhibition (OFC)},
  pages={1--3},
  year={2025},
  organization={IEEE}
}

@article{chiti2025_digital_twin_sdn_qn,
  title={Towards Digital-Twin Assisted Software-Defined Quantum Satellite Networks},
  author={Chiti, Francesco and Pecorella, Tommaso and Picchi, Roberto and Pierucci, Laura},
  journal={Sensors (Basel, Switzerland)},
  volume={25},
  number={3},
  pages={889},
  year={2025}
}

@inproceedings{luo2025_digital_twin_scalable_cloud,
  title={A Digital Twin of Scalable Quantum Clouds},
  author={Luo, Waylon and Baheri, Betis and Humble, Travis and Zhao, Jiapeng and Zhan, Tong and Maharjan, Rajan and Guan, Qiang},
  booktitle={39th ACM SIGSIM Conference on Principles of Advanced Discrete Simulation},
  pages={165--175},
  year={2025}
}

@article{li2021_fpga_emulation,
  title={FPGA-accelerated quantum computing emulation and quantum key distillation},
  author={Li, He and Pang, Yaru},
  journal={IEEE Micro},
  volume={41},
  number={4},
  pages={49--57},
  year={2021},
  publisher={IEEE}
}

@inproceedings{chung2025_cross_validation,
  title={Cross-validating quantum network simulators},
  author={Chung, Joaquin and Hajdu{\v{s}}ek, Michal and Benchasattabuse, Naphan and Kolar, Alexander and Singal, Ansh and Soon, Kento Samuel and Teramoto, Kentaro and Zang, Allen and Kettimuthu, Raj and Van Meter, Rodney},
  booktitle={IEEE INFOCOM 2025-IEEE Conference on Computer Communications Workshops (INFOCOM WKSHPS)},
  pages={1--6},
  year={2025},
  organization={IEEE}
}

@article{chalupnik2025_realistic_bbm92_simulation,
  title={Realistic quantum network simulation for experimental BBM92 key distribution},
  author={Chalupnik, Michelle and Doolittle, Brian and Seshadri, Suparna and Brown, Eric G and Kenemer, Keith and Winton, Daniel and Sanchez-Rosales, Daniel and Skrzypczyk, Matthew and Alexander, Cara and Ostby, Eric and others},
  journal={arXiv preprint arXiv:2505.24851},
  year={2025}
}

@inproceedings{khalid2004fpga,
  title={FPGA emulation of quantum circuits},
  author={Khalid, Ahmed Usman and Zilic, Zeljko and Radecka, Katarzyna},
  booktitle={IEEE International Conference on Computer Design: VLSI in Computers and Processors, 2004. ICCD 2004. Proceedings.},
  pages={310--315},
  year={2004},
  organization={IEEE}
}

@article{mahmud2019_scaling_qc_emulation,
  title={Scaling reconfigurable emulation of quantum algorithms at high precision and high throughput},
  author={Mahmud, Naveed and El-Araby, Esam and Caliga, David},
  journal={Quantum Engineering},
  volume={1},
  number={2},
  pages={e19},
  year={2019},
  publisher={Wiley Online Library}
}

@article{el2023_qc_classical_emulation,
  title={Towards complete and scalable emulation of quantum algorithms on high-performance reconfigurable computers},
  author={El-Araby, Esam and Mahmud, Naveed and Jeng, Mingyoung Jessica and MacGillivray, Andrew and Chaudhary, Manu and Nobel, Md Alvir Islam and Islam, SM Ishraq Ul and Levy, David and Kneidel, Dylan and Watson, Madeline R and others},
  journal={IEEE Transactions on Computers},
  volume={72},
  number={8},
  pages={2350--2364},
  year={2023},
  publisher={IEEE}
}

@article{mourya2023_qc_emulation,
  title={Emulation of quantum algorithms using CMOS Analog Circuits},
  author={Mourya, Sharan and La Cour, Brian R and Sahoo, Bibhu Datta},
  journal={IEEE Transactions on Quantum Engineering},
  volume={4},
  pages={1--16},
  year={2023},
  publisher={IEEE}
}

@inproceedings{zhang2015_gpu,
  title={Quantum computer simulation on multi-GPU incorporating data locality},
  author={Zhang, Pei and Yuan, Jiabin and Lu, Xiangwen},
  booktitle={International Conference on Algorithms and Architectures for Parallel Processing},
  pages={241--256},
  year={2015},
  organization={Springer}
}

@inproceedings{doi2019_qc_sim_hpc,
  title={Quantum computing simulator on a heterogenous HPC system},
  author={Doi, Jun and Takahashi, Hitomi and Raymond, Rudy and Imamichi, Takashi and Horii, Hiroshi},
  booktitle={Proceedings of the 16th ACM International Conference on Computing Frontiers},
  pages={85--93},
  year={2019}
}

@article{horii2023_gpu,
  title={Efficient techniques to gpu accelerations of multi-shot quantum computing simulations},
  author={Horii, Hiroshi and Wood, Christopher and others},
  journal={arXiv preprint arXiv:2308.03399},
  year={2023}
}

@inproceedings{faj2023_gpu,
  title={Quantum computer simulations at warp speed: Assessing the impact of gpu acceleration: A case study with ibm qiskit aer, nvidia thrust \& cuquantum},
  author={Faj, Jennifer and Peng, Ivy and Wahlgren, Jacob and Markidis, Stefano},
  booktitle={2023 IEEE 19th International Conference on e-Science (e-Science)},
  pages={1--10},
  year={2023},
  organization={IEEE}
}

@article{Ladd2010_qc,
  title = {Quantum computers},
  volume = {464},
  ISSN = {1476-4687},
  url = {http://dx.doi.org/10.1038/nature08812},
  DOI = {10.1038/nature08812},
  number = {7285},
  journal = {Nature},
  publisher = {Springer Science and Business Media LLC},
  author = {Ladd,  T. D. and Jelezko,  F. and Laflamme,  R. and Nakamura,  Y. and Monroe,  C. and O’Brien,  J. L.},
  year = {2010},
  month = mar,
  pages = {45–53}
}

@misc{resch2019_cq_comparison,
      title={Quantum Computing: An Overview Across the System Stack}, 
      author={Salonik Resch and Ulya R. Karpuzcu},
      year={2019},
      eprint={1905.07240},
      archivePrefix={arXiv},
      primaryClass={quant-ph},
      url={https://arxiv.org/abs/1905.07240}, 
}

@article{lloyd1996universal,
  title={Universal quantum simulators},
  author={Lloyd, Seth},
  journal={Science},
  volume={273},
  number={5278},
  pages={1073--1078},
  year={1996},
  publisher={American Association for the Advancement of Science}
}

@article{georgescu2014_quantum_simualation_review,
  title={Quantum simulation},
  author={Georgescu, Iulia M and Ashhab, Sahel and Nori, Franco},
  journal={Reviews of Modern Physics},
  volume={86},
  number={1},
  pages={153--185},
  year={2014},
  publisher={APS}
}

@article{altman2021_quantum_sim,
  title={Quantum simulators: Architectures and opportunities},
  author={Altman, Ehud and Brown, Kenneth R and Carleo, Giuseppe and Carr, Lincoln D and Demler, Eugene and Chin, Cheng and DeMarco, Brian and Economou, Sophia E and Eriksson, Mark A and Fu, Kai-Mei C and others},
  journal={PRX quantum},
  volume={2},
  number={1},
  pages={017003},
  year={2021},
  publisher={APS}
}

@article{roffe2019_qec,
  title={Quantum error correction: an introductory guide},
  author={Roffe, Joschka},
  journal={Contemporary Physics},
  volume={60},
  number={3},
  pages={226--245},
  year={2019},
  publisher={Taylor \& Francis}
}

@article{cai2023_error_mitigation,
  title={Quantum error mitigation},
  author={Cai, Zhenyu and Babbush, Ryan and Benjamin, Simon C and Endo, Suguru and Huggins, William J and Li, Ying and McClean, Jarrod R and O’Brien, Thomas E},
  journal={Reviews of Modern Physics},
  volume={95},
  number={4},
  pages={045005},
  year={2023},
  publisher={APS}
}

@article{avis2025_repeater_placement,
  title={Optimization of quantum-repeater networks using stochastic automatic differentiation},
  author={Avis, Guus and Krastanov, Stefan},
  journal={Physical Review Research},
  volume={7},
  number={3},
  pages={033111},
  year={2025},
  publisher={APS}
}

@article{chen2023inferring_qn_topology,
  title={Inferring quantum network topology using local measurements},
  author={Chen, Daniel T and Doolittle, Brian and Larson, Jeffrey and Saleem, Zain H and Chitambar, Eric},
  journal={PRX Quantum},
  volume={4},
  number={4},
  pages={040347},
  year={2023},
  publisher={APS}
}

@article{doolittle2023_vqo,
  title={Variational quantum optimization of nonlocality in noisy quantum networks},
  author={Doolittle, Brian and Bromley, R Thomas and Killoran, Nathan and Chitambar, Eric},
  journal={IEEE Transactions on Quantum Engineering},
  volume={4},
  pages={1--27},
  year={2023},
  publisher={IEEE}
}

@article{doolittle2024_operational_framework,
  title={An Operational Framework for Nonclassicality in Quantum Communication Networks},
  author={Doolittle, Brian and Leditzky, Felix and Chitambar, Eric},
  journal={arXiv preprint arXiv:2403.02988},
  year={2024}
}

@article{riera2025_qn_simulation,
  title={Quantum Simulation of Noisy Quantum Networks},
  author={Riera-S{\`a}bat, Ferran and Miguel-Ramiro, Jorge and D{\"u}r, Wolfgang},
  journal={arXiv preprint arXiv:2506.09144},
  year={2025}
}

@article{peng2020_quantum_circuit_cutting,
  title = {Simulating Large Quantum Circuits on a Small Quantum Computer},
  author = {Peng, Tianyi and Harrow, Aram W. and Ozols, Maris and Wu, Xiaodi},
  journal = {Phys. Rev. Lett.},
  volume = {125},
  issue = {15},
  pages = {150504},
  numpages = {6},
  year = {2020},
  month = {Oct},
  publisher = {American Physical Society},
  doi = {10.1103/PhysRevLett.125.150504},
  url = {https://link.aps.org/doi/10.1103/PhysRevLett.125.150504}
}

@article{harada2024_doubly_optimal_wire_cutting,
  title = {Doubly Optimal Parallel Wire Cutting without Ancilla Qubits},
  author = {Harada, Hiroyuki and Wada, Kaito and Yamamoto, Naoki},
  journal = {PRX Quantum},
  volume = {5},
  issue = {4},
  pages = {040308},
  numpages = {43},
  year = {2024},
  month = {Oct},
  publisher = {American Physical Society},
  doi = {10.1103/PRXQuantum.5.040308},
  url = {https://link.aps.org/doi/10.1103/PRXQuantum.5.040308}
}

@ARTICLE{martin2022_quantum_datacenter,
  author={Martin, Michael James and Hughes, Caroline and Moreno, Gilberto and Jones, Eric B. and Sickinger, David and Narumanchi, Sreekant and Grout, Ray},
  journal={IEEE Transactions on Sustainable Computing}, 
  title={Energy Use in Quantum Data Centers: Scaling the Impact of Computer Architecture, Qubit Performance, Size, and Thermal Parameters}, 
  year={2022},
  volume={7},
  number={4},
  pages={864-874},
  doi={10.1109/TSUSC.2022.3190242}}

@article{shapourian2025_quantum_datacenter,
  title={Quantum data center infrastructures: A scalable architectural design perspective},
  author={Shapourian, Hassan and Kaur, Eneet and Sewell, Troy and Zhao, Jiapeng and Kilzer, Michael and Kompella, Ramana and Nejabati, Reza},
  journal={arXiv preprint arXiv:2501.05598},
  year={2025}
}

@inproceedings{haner2016_qc_emulator,
  title={High performance emulation of quantum circuits},
  author={H{\"a}ner, Thomas and Steiger, Damian S and Smelyanskiy, Mikhail and Troyer, Matthias},
  booktitle={SC'16: Proceedings of the International Conference for High Performance Computing, Networking, Storage and Analysis},
  pages={866--874},
  year={2016},
  organization={IEEE}
}

@article{elyasi2025framework,
  title={A Framework for Quantum Data Center Emulation Using Digital Quantum Computers},
  author={Elyasi, Seyed Navid and Ahmadian, Seyed Morteza and Monti, Paolo and Li, Jun and Lin, Rui},
  journal={arXiv preprint arXiv:2509.04029},
  year={2025}
}

@inproceedings{kefaloukos2025internet,
  title={The Internet of Quantum Things (IoQT)-A New Frontier in Quantum Emulation and Simulation},
  author={Kefaloukos, Ioannis and Tcholtchev, Nikolay and Kourtis, Michail-Alexandros and Oikonomakis, Giorgos and Rompogiannakis, Emmanouil Eleftherios and Markakis, Evangelos},
  booktitle={Joint National Conference on Cybersecurity 2025},
  year={2025}
}

@article{lauk2020_transduction,
  title={Perspectives on quantum transduction},
  author={Lauk, Nikolai and Sinclair, Neil and Barzanjeh, Shabir and Covey, Jacob P and Saffman, Mark and Spiropulu, Maria and Simon, Christoph},
  journal={Quantum Science and Technology},
  volume={5},
  number={2},
  pages={020501},
  year={2020},
  publisher={IOP Publishing}
}

@inproceedings{amir2022_quantum_twins,
  title={What can we expect from Quantum (Digital) Twins?},
  author={Amir, Malik and Bauckhage, Christian and Chircu, Alina and Czarnecki, Christian and Knopf, Christian and Piatkowski, Nico and Sultanow, Eldar},
  booktitle={17th International Conference on Wirtschaftsinformatik},
  year={2022}
}

@article{Supic2020_self-testing,
  doi = {10.22331/q-2020-09-30-337},
  url = {https://doi.org/10.22331/q-2020-09-30-337},
  title = {Self-testing of quantum systems: a review},
  author = {{\v{S}}upi{\'{c}}, Ivan and Bowles, Joseph},
  journal = {{Quantum}},
  issn = {2521-327X},
  publisher = {{Verein zur F{\"{o}}rderung des Open Access Publizierens in den Quantenwissenschaften}},
  volume = {4},
  pages = {337},
  month = sep,
  year = {2020}
}

\end{document}